\documentclass[pra,onecolumn,tightenlines]{revtex4}

\usepackage[english]{babel}
\usepackage{graphicx}
\usepackage{bm}
\usepackage{color}

\begin{document}

\title{Single-photon device requirements for operating linear optics quantum computing outside the post-selection basis}

\author{Thomas Jennewein$^{a,b,c}$$^{\ast}$\thanks{$^\ast$Corresponding author. Email: thomas.jennewein@uwaterloo.ca}
Marco Barbieri$^{b,d}$ and 
Andrew G. White$^{b}$ \\ 
$^{a}${\em{Institute for Quantum Computing, University Waterloo, Waterloo ONT, Canada}};
$^{b}${\em{Department of Physics, and Centre for Quantum Computer Technology, the University of Queensland, 4027 St Lucia, QLD, Australia}};
$^{c}${\em{Institute for Quantum Information and Quantum Optics, Austrian Academy of Sciences, Vienna, Austria}};
$^{d}${\em{Laboratoire Charles Fabry, Institut d'Optique, 91127 Palaiseau, France}};}


%

\begin{abstract}
Photonics is a promising architecture for the realisation of quantum information processing, since the two-photon interaction, or non-linearity, necessary to build logical gates can efficiently be realised by the use of interference with ancillary photons and detection.\cite{Knill:2001lr}  Although single-photon sources and detectors are pivotal in realisations of such systems, clear guidelines for the required performance of realistic systems are yet to be defined. We present our detailed numerical simulation of several quantum optics circuits including sources and detectors all represented in multi-dimensional Fock-spaces, which allows to obtain experimentally realistic performance bounds for for these devices. In addition, the single-photon source based on switched parametric down-conversion is studied, which in principle could reach the required performance. Three approaches for implementing the switching hierarchy of the photons are simulated, and their anticipated performance is obtained. Our results define the bar for the optical devices needed to achieve the first level of linear-optics quantum computing outside the coincidence basis.
\end{abstract}

\maketitle

%

\section{Introduction}

With scalable linear optical quantum computing being the long term goal, even the obvious near term experimental steps such as heralded quantum gates or creation of entangled states are very interesting applications for quantum communication, metrology and information processing.   The performance of linear optical quantum computing (LOQC) relies on the quality of the various elements and devices that the system is built from, such as sources, optical elements and detectors. 

The main concepts of LOQC where discovered by Knill, Milburn and Laflamme (KLM) in 2001\cite{Knill:2001lr}, {\color{black} allowing} the near-deterministic {\color{black} performance} of a scalable two-photon gate (e.g. controlled not). Within a short time LOQC was further advanced with the discovery of one-way quantum computing \cite{raussendorf2001}, where the main difficulty is creating the cluster state. The original idea to create such a state is to perform KLM-type scalable gates between several photons. But later, more efficient methods for creating the cluster states have also been discovered, such as using fusion gates \cite{browne-2005-95}. 

All these linear optical quantum computing circuits are based on the assumption of ideal elements,  that is perfect efficiency of optics,  {\color{black} highly} efficient and {\color{black} photon-number} resolving detectors, efficiency ${\color{black}\approx} 1$, and perfect single photon sources, {\color{black} output probability $\approx 1$}. 
{\color{black} Even though it has recently be shown that in the asymptotic limit LOQC could work if the product of the detector efficiency and the single photon output probability is greater than $2/3$  \cite{PhysRevLett.100.060502} , (or even greater than $1/2$ in the case of pair sources \cite{PhysRevA.81.052303}), these striking results are not too practical as they require photon sources that do not emit any higher number photon numbers or multi-photon states, and very broad and complex optical circuits. } 
A further assumption required for achieving ideal photon interference quality is that the photon  {\color{black} wave-packets} are Fourier limited and spectrally pure, and that the mode matching of the beams is ideal. Unfortunately, realistic systems will never allow us to reach the ideal case. Indeed, there are known levels of errors for performing the full scale quantum computing, which however are hard to achieve in the laboratory. 

In this work we revisit the single-photon devices required for implementing basic linear optics quantum computing (LOQC) circuits 
{\color{black} in view of achieving the next experimental level, i.e. performing the operations outside the post-selection basis.}
 Post-selection is commonly implemented in photon-experiments to cope with the inefficiencies of devices such as optical losses by only {\color{black} considering }
 successful experimental runs, {\color{black} where the desired number of photons is actually observed}. In current experimental implementations, the optical efficiencies are in the order of 10\% and source efficiencies around 1\%, leaving the ratio of successful to unsuccessful attempts below $10^{-6}$. Therefore, current LOQC experiments involve only up to six photons, and require extensive post-selection and long measurement times.  If such circuits could be performed without post-selection, quantum processing with multiple-photon would gain significant grounds and much larger numbers of photonic qubits would become accessible. 

 In order to understand just how good the single-photon devices must perform to reach this important goal, we studied their parameters through numerical simulation of the envisioned quantum optical circuits{\color{black}; we included} the crucial deteriorating effects such as multi-photon emissions, optical losses and non-ideal detection. To some extent, the source and detector requirements have been addressed for measurement-based quantum computing in\cite{browne-2005-95}, however their analysis included huge numbers of sources and detectors, and some idealizations. Our simulation is experiment-driven and it aims to obtain and study accessible parameters and somewhat feasible LOQC implementations to characterize the sources and detectors.

\section{Numerical simulation of quantum optics}

The important feature of our approach was that every photon mode was treated as a $N$-dimensional Fock-space, {\color{black}accounting for several higher-number photon terms}. Accordingly, all operators were implemented as matrices of dimension $N \times N$ obtained by matrix exponentiation of the interaction Hamiltonians, requiring significant computational resources for multi-mode optical circuits. 
	
{\color{black} The numerical model was implemented in the {\em Matlab} environment, based on the quantum optics toolbox by Sze M. Tan\cite{1464-4266-1-4-312}. A collection of functions and scripts were developed and specifically designed for studying single photon experiments \cite{Jennewein2010}.} In this model, every photon mode is represented by a Fock-space of the size N. Each state is a vector, where the first element corresponds to the amplitude of the vacum state $|0\rangle$, and the last element corresponds to the amplitude of $N-1$ photon state, $|n-1\rangle$. It is straightforward to represent a photonic qubit, such as the polarization of photons, as a tensor product of two such Fock-space modes.  All operations on the photon modes must be realized by an unitary evolution matrix, generated from exponentiation of the corresponding Hamiltonian. In order to explain how the representation of quantum states and operators {\color{black} are} incorporated we have given some examples in the Appendix.

\subsection{Time discretization}

This model is  a monochromatic approximation of our system, therefore does not accurately incorporate any time evolution. Even though this might seem a strong approximation, it is sufficient for general simulation and analysis of LOQC type experiments.  In a typical multi-photon experiment \cite{Bouwmeester97a} the photon wave-packet is as short as 500 femto seconds, whereas the detector electronic timing resolution is on the order of 500 pico seconds. Hence, the detectors are essentially integrating the optical field over their timing resolution. As outlined in \cite{Zukowski95a}, it is possible to treat the entangling operations, such as two-photon interference, occuring between the various photons by only observing the overall behaviour of the detectable events without needing to describe and model the concrete and microscopic time evolution, as long as the generation times of the various photon are synchronised to a time \emph{better} than the photon time envelopes, which is achieved by ultra-short laser pulses.  

In our numerical model we assume this condition is satisfied, and therefore the calculations are performed in a time-unit $\tau_d$ corresponding to the detector timing resolution.

\subsection{Photon source}

The output of the realistic photon sources are modeled with terms for single-photons $\epsilon$ , and also for the two-photons $\omega$, as follows:
\begin{equation}
 |\zeta\rangle = M (\sqrt{1-\epsilon^2}|0\rangle + \epsilon\sqrt{1-\omega^2}|1\rangle + \epsilon\omega |2\rangle),
 \label{source_model}
 \end{equation}
with  normalization $M$.  The single-photon performance of the source will be characterized via the second order correlation function at zero time delay, defined as $g^{(2)}(0) = \langle {a^\dagger }^2 a^2 \rangle/  \langle a^\dagger a\rangle^2$. The output success probability of a source is the probability of finding any photons in the signal, given be the inverse of finding an empty output $P_{hrld} = 1 -  \langle0|\zeta\rangle \langle \zeta   |0\rangle$. This description of the single photon source, Eq. \ref{source_model}, has the nice feature, that the correlation value $g^{(2)}(0)$ is essentially constant over $P_{hrld} $. 
{\color{black} In particular, as this correlation value is an easily measured quality estimator  it is widely used in experimental implementations of single-photon sources, and it will allow the direct relation of our simulation results to such implementations. }

As {\color{black}one} realistic implementation of photon sources we considered heralding photons from parametric down-conversion (SPDC). This will be discussed in full detail in section~\ref{herald_spdc}.

\subsection{Single-photon detectors}

\subsubsection{Bucket Detector - BD}
The first type is the ``bucket detector", which will only give a ``click"  in the presence of photons, but will not discriminate between photon numbers. A typical example for such a detector is an avalanche photo diode (APD). The probability for a click is calculated with a projection operator, where the diagonal terms of the projector matrix take the values 
\begin{equation}
P_{BD}(i,i) = 1-(1-\eta)^{(i-1)} \label{proj_apd}, 
\end{equation}
where $i$ corresponds to the particular photon number plus one, and the total efficiency $\eta$ is the product of the actual detector efficiency $\eta_{BD}$ with the optical efficiency $\eta_{optics}$ to include losses from optical coupling and the system.  

Also of interest is the {\color{black}operator describing an unsuccessful detection event, ``no click'',} which has the relevant applications for the suppression of higher-order terms in two-channel polarisation analysers. The corresponding projector based on the same efficiency as above, has the diagonal elements: 
\begin{equation}
\bar{P}_{BD}(i,i) = (1-\eta)^{(i-1)}.
\end{equation}

{\color{black} These operators for successful detection and non-detection of photons are the important tools required to simulate the evolution of the quantum states and expectation values of measurements.}

\subsubsection{Single photon sensitive detector - SPD}

The more important type of photon detector for LOQC has the ability to discriminate photon numbers. Here we focus on the detection of one-and-only-one photon, hence named a single photon detector (SPD). These are modelled as an operator matrix with the diagonal elements according to the probability of obtaining a positive detection signal for a given input photon state, which for a given total efficiency $\eta$ has the following diagonal terms:
\begin{equation}
P_{SPD}(i,i) = (i-1)\eta(1-\eta)^{(i-2)},  \label{proj_spd}
\end{equation}
as well as the  the No-Click-projector
\begin{equation}
\bar{P}_{SPD}(i,i) = (1-\eta)^{(i-1)},
\end{equation}
where as above, the efficiency is $\eta = \eta_{SPD} * \eta_{optics}$,  the product of the intrinsic detector efficiency and the optical system, and $i$ ranges from $1...N$.

\subsubsection{Noise and saturation of the photon detectors}
In addition to the main deficiency of detectors which is the limited efficiency, they also show error detections due to dark counts. This describes the behaviour of detectors to give false detection events, even if no signal is present. We model this by adding a constant term $\epsilon_d$, corresponding to the probability of a noise induced detection signal per time unit  $\tau_d$, to each element of the detectors operator:
\begin{equation}
P_{ndet}(i,i) = P(i,i)_{BD/SPD} + \epsilon_d . 
\end{equation}
Generally, one must take care that the final calculated probability for a detection will in general not be continuos, but rather must also include a maximal cutoff at unity which actually  corresponds  to saturation of the detector. The probability for a detection event then would be limited to unity in the following relation $P=\max (Tr(P_{ndet} | \Phi \rangle),1)$. It is straightforward to see why this should be correct: in the case of a unit efficiency detector with a dark count probability $\epsilon_d$, the final detection probability of a signal will be offset by this amount, until the level of unity is reached and the detector saturates. However, in all practical cases that are considered here, the detection probability per time-unit is significantly less than unity, and therefore this cutoff must not be implemented. 

In the case of the No-Click projector we subtract $\epsilon_d$ from each term, however it must be ascertained that {\color{black}none} of the values becomes negative. More practically in actual experiments, the dark count rate can be efficiently decreased by an appropriate time gating, triggered, for instance, by the main pump laser.

%

%

%
\section{Simulation of LOQC circuits and the results}

With the numeric modelling of quantum optics in place we were able to conduct a series of simulations on interesting linear optical quantum circuits. The chosen small selection is representative for both the existing experiments - non-scalable gates that operate in post-selection -  as well as for the next steps towards future implementations of LOQC - scalable gates, that do not require post-selection. 
{\color{black} What is unique in our approach is that we aimed to follow the practical experimental implementation for the studied quantum circuit, as well as the actual methods for generating single photons from cascaded down-conversion and non-perfect single-photon detectors. }

\subsection {C-Phase gate - non-scalable scheme 1: } We simulate the  creation of Bell-pairs of photons (See Figure \ref{simulated_schemes_non_scalable}, a) which is easily achieved by {\color{black} making two photons interfere} on a partial polarising beam splitter (PPBS) which acts as a controlled-phase  (c-phase) gate \cite{PhysRevLett.95.210504}. This is non-scalable operation, as it can only work in post-selection meaning that the desired photon state is extracted by detection of the photons. This scheme can in general not be concatenated to implement more complex operations, except in few very special cases {\color{black}as illustrated in the next section}. {\color{black} The quality of the output is assessed} via the violation of the Bell-pair witness\cite{PhysRevLett.91.227901}, as well as the Clauser-Horne-Shimony-Holt-inequality (CHSH-inequality) violation\cite{PhysRevLett.23.880}. 

We find from the simulation that in this case the involved single photon sources must show a $g^2 (0)$-correlation less than 0.15 in order that a Bell-inequality can be violated, and less than 0.39 in order to satisfy the Bell-state witness\cite{Guhne:2003fk}. 
{\color{black} Due to postselection, the actual detection efficiency and the success probability of the setup are not fundamental issues as they only affect the achievable count rate but not the quality of the output.}

\subsection{Two C-Phase gates - non-scalable scheme 2:} 
\begin{figure}[htbp]
\begin{center}
\includegraphics[width=9.5cm]{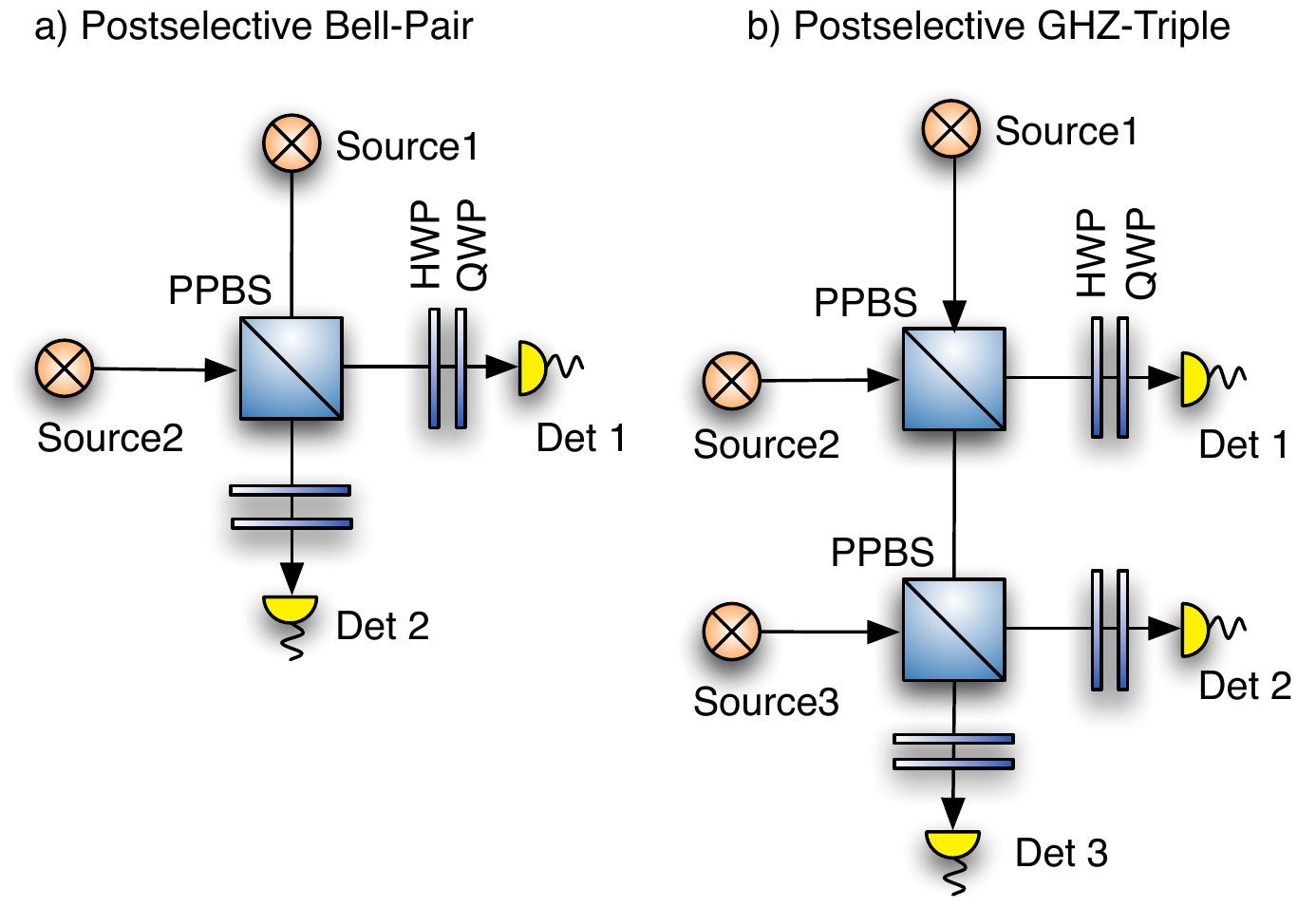}\\
\caption{Schematics of the two post-selective (and therefore non-scalable)  quantum circuits that were studied here.  a) is a post-selective implementation of a c-phase gate, used to create a Bell-pair of photons, b) scheme of concatenated c-phase gates for creating Greenberger-Horne-Zeilinger (GHZ)-triples of photons.}
\label{simulated_schemes_non_scalable}
\end{center}
\end{figure}

By chaining two of the above mentioned non-scalable c-phase gates (Figure \ref{simulated_schemes_non_scalable}, b), it is possible to entangle three photons into a Greenberger-Horne-Zeilinger-state (GHZ). As was the case of just one c-phase gate, this circuit can only work in post-selection and is therefore not scalable. 
{\color{black} We reconstruct the value of the witness for a GHZ state\cite{Guhne:2003fk} from our simulation as the quantitative criterion to assess the performance of the entangling operation.}

{\color{black} Our simulation shows} that the required quality of the single-photons must be to proved an output with a $g^2 (0)$ correlation less than 0.06, which is already a huge challenge to achieve. 
{\color{black} Also in this case, the efficiency of the detectors and of the source only affects the count rate. }

\subsection{ KLM-type gates - scalable scheme 1:} In order to achieve bounds for the next important step towards LOQC, we simulated the heralding of Bell-pairs with KLM-type gates. Thereby the control and target photons interact via the assistance (effectively two-photon interferences) of  two ancillary photons. The final detection of the ancillary photons {\color{black}flags a successful run of the gate}, and the control and target photon are retrievable - and reusable - from their outputs. These gates are fully scalable and can {\color{black}build} arbitrary circuits. It is important to note that these gates require single-photon resolving detectors. 
{\color{black}We focussed our simulation in particular on the simplified KLM-gate\cite{PhysRevA.68.042328} , (Figure \ref{simulated_schemes_scalable}, left). We also studied the cases of the Knill-gate \cite{PhysRevA.66.052306} and the Pittman-Franson-gate \cite{PhysRevA.64.062311}. We observed that the former has performances nearly identical to the KLM gate. Concerning the latter, this is harder to implement, as it requires a Bell pair as an ancilla; this demands either a deterministic entanglement source or a KLM gate on its own. }
In order to test the operation of the gate we chose to follow exactly what a possible experimental situation would be in order to prove that this gate was successfully realised: we observed the pair-wise correlations of the output photons for the Clauser-Horne (CH) inequality\cite{Clauser74a}, which explicitly {\color{black}excludes post-selection in the measurements or analysis.} In addition, we also studied the witness for a Bell-state. 

\begin{figure*}[htbp]
\begin{center}

\includegraphics[width=12cm]{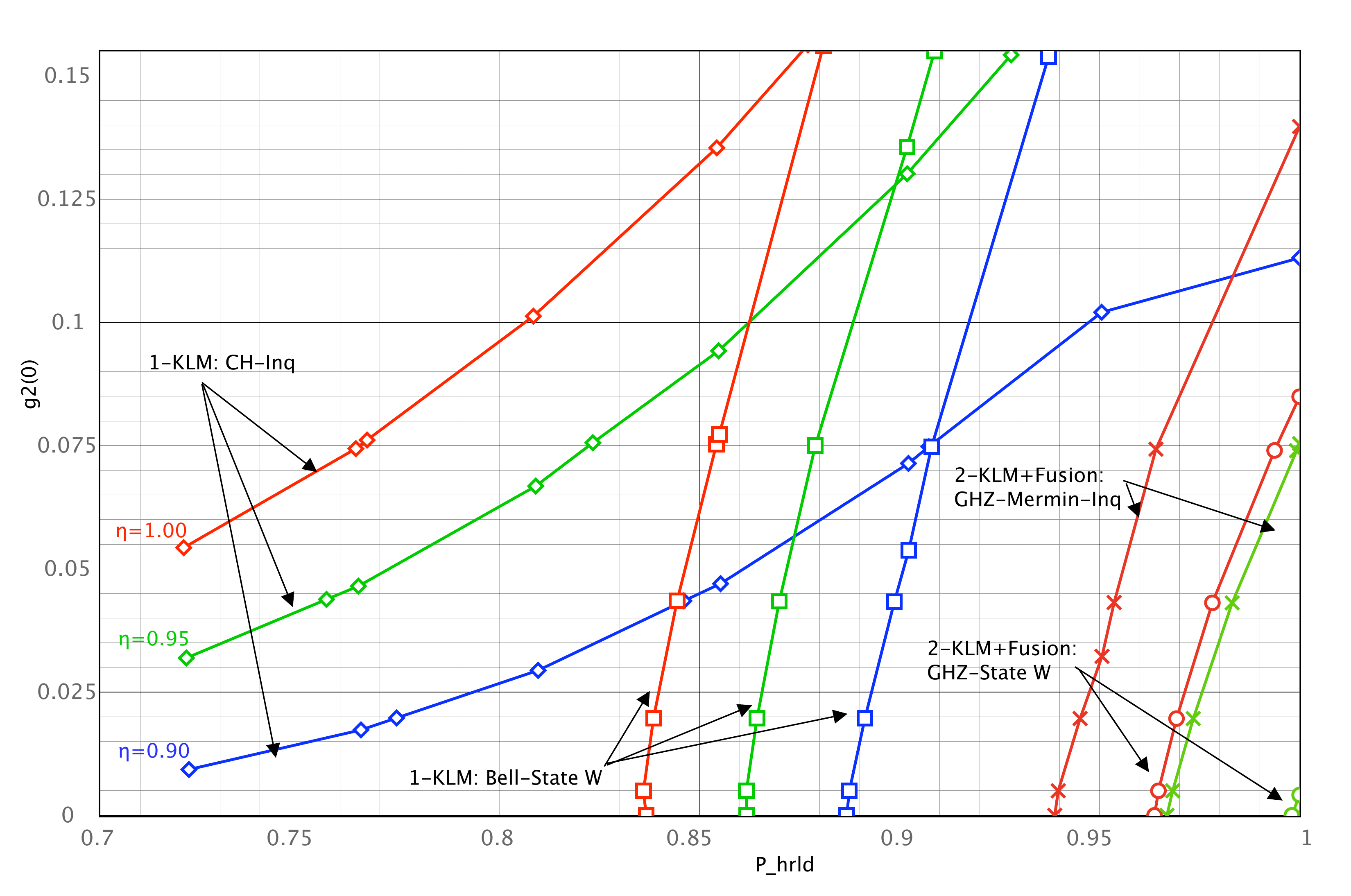}

\caption{Numerical results of source and detector performance requiblack to achieve performing simple linear optical quantum computing ( LOQC) circuits without post-selection.  1-KLM denotes the calculation where two photons are entangled into a Bell-states via one simplified Knill-Laflamme-Milburn gate; 2-KLM denotes the operation where two such Bell-states are fused, and a 3-photon GHZ state is created. The vertical axis is the second order correlation of the photon source beams, $g^2 (0)$ at zero time delay. The horizontal axis is the probability that  the involved photon sources does fire upon a trigger and herald a photon signal, $P_{hrld}$. The simulations where performed assuming total efficiencies (optical and detectors) between $\eta= 0.90, 0.95, 1.00$, coded in blue, green and black. The simulation showed that the total efficiency must be greater than about {\color{black}$\eta=0.9$} to deliver useful results in the 1-KLM case, and above $\eta = 0.95$ for the 2-KLM case.}
\label{results}
\end{center}
\end{figure*}

A good way of presenting the photon-source requirements is to  plot of their performance in terms of {\color{black}$g^{(2)} (0)$}-correlation as well as the success probability for heralding a single photon, see Figure~\ref{results}, 1-KLM plots. Summarising all the simulation results becomes very  {\color{black}cumbersome}, as many parameters in the simulation are varied, and extracting the actual interesting information is hard for the reader.  Here, as can be seen in the figure, the total system efficiency (i.e. optics + detectors) becomes crucial. Only at the total system efficiency above $\eta=0.9$ and single-photon resolving detection {\color{black} the simplified-KLM-gate performs in a useful manner.} In this case, to give concrete numbers, if the probability for the source to herald single photons is 0.8  then the requiblack $g^2 (0)$-correlation must be below 0.03 in order to violate the CH-inequality. Notably, satisfying the Bell-state Witness can only be achieved if the probability for the source to herald single photons is above  0.9, then the $g^2 (0)$-correlation may become as high as 0.045. The source requirements do become relaxed if the system-detection efficiency is higher than 0.9.

\subsection {One-way quantum computing - scalable scheme 2:} 

\begin{figure*}[htbp]
\begin{center}

\includegraphics[width=7.5cm]{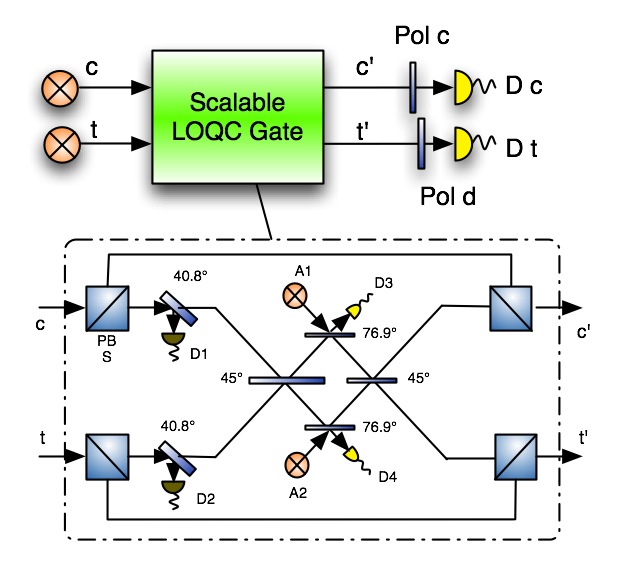}
\includegraphics[width=8cm]{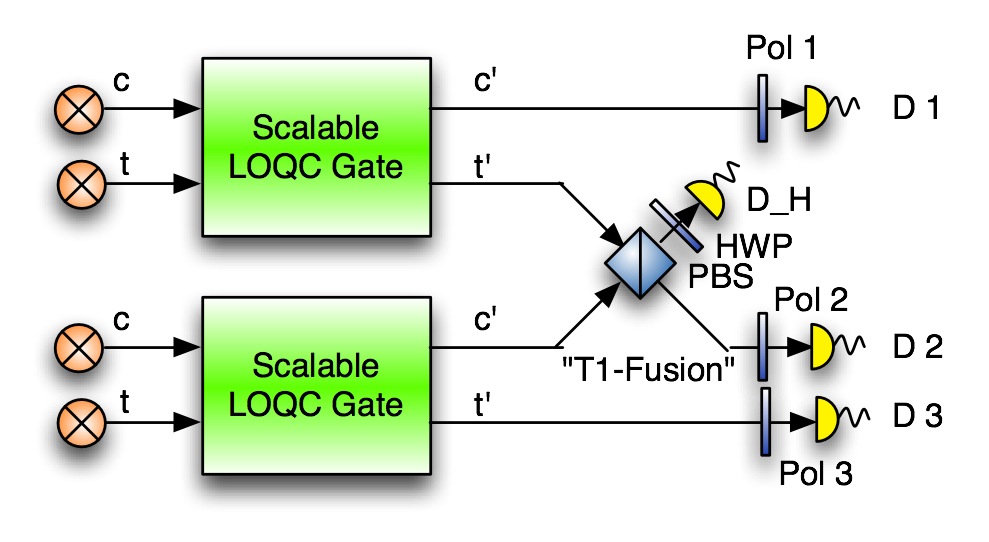}

\caption{Schematics of the simulated scalable-LOQC circuits.  (left) Scalable linear optical gate called the simplified-KLM-gate (inset), utilised to create heralded Bell-pairs of photons form the initially separable input photons. The successful operation of this gate is flagged by the detection of the two ancillary single photons in their outputs. The quality of the gate is observed by testing the CH-Bell-inequality on the entangled photons which does not allow for any post-selection on the output states. (right) As the first level of the one-way quantum computing with cluster states the direct and heralded generating of a three-photon GHZ state is studied. GHZ-states are created from two heralded Bell-pairs generated in the simplyfied-KLM-gate, connected via a type-I Fusion gate.}
\label{simulated_schemes_scalable}
\end{center}
\end{figure*}

Cluster-state quantum computing can be efficiently realised if a source {\color{black}of} heralded GHZ-entangled photons is available. There are many approaches to directly creating such states, however LOQC itself offers the possibility to generate a heralded GHZ states by fusing two Bell-pairs created from separate KLM-gates, as outlined in \cite{browne-2005-95}, (Figure \ref{simulated_schemes_scalable}, right). The fusion-operation is in interference of two-photons on a beam-splitter and detection of exactly one photon in one output arm. The resulting GHZ-entangled state of the three photons is tested against a Mermin-type inequality\cite{Mermin90a} and a GHZ-state witness, \cite{Guhne:2003fk}. 

The simulation of this scenario became very resource-hungry, since the total amount of modes is rather large. The main outcomes from the simulations clearly show that this scenario will also be very challenging to realise. The total detection efficiency $\eta$ should be at lest $0.95$, and the requirements for the photon-sources are tight, asking for a probability for heralding output photons $P_{hrld} > 0.97$ and the $g^2 (0)$-correlation of $0.01$.  Reaching such requirements will be very challenging for any optical systems.

\section{Generating single photons} \label{herald_spdc}

These simulations have provided the baseline for the requirements that such LOQC circuits pose on photon-sources and photon-detectors. In the following it will be studied, if, and how, a single-photon source based on parametric down-conversion might be able to reach these challenging parameters.

The presently most important source for multiple and entangled photons is spontaneous parametric down conversion\cite{N.:1967kb}, which is essentially the spontaneous emission from an optical parametric amplifier, also referblack to as squeezed vacuum. When observing the two output modes of the squeezer, photons are produced in pairs, and therefore the correlations are sub-poissonian, and therefore enable doing conditional single photon experiments. However, given the non-deterministic  behaviour of this interaction, the  photon-state created from SPDC produces not only single photon-pair terms, but also multiple numbers of photon-pairs.

\subsection{Modelling of SPDC}
The Hamiltonian that describes the interaction in SPDC is the simplified squeezing operator, which has the approximation of a strong pump mode, and acts on two output modes $a_1,a_2$ in the form
\begin{equation}
H_{SPDC}=\epsilon(a_1^\dagger a_2^\dagger + a_1 a_2), \label{H_spcd}
\end{equation}
where $\epsilon$ is the squeezing parameter.
The evolution of this Hamiltonian calculated via matrix exponentiation. 
Obviously, our numerical model takes account of all higher order photon number terms with their full and correct amplitude which can then easily be propagated through any quantum circuits simulation, such as logic gates.

\subsection{Heralding Photons from SPDC}
\begin{figure*}[htbp]
\begin{center}

\includegraphics[width=13cm]{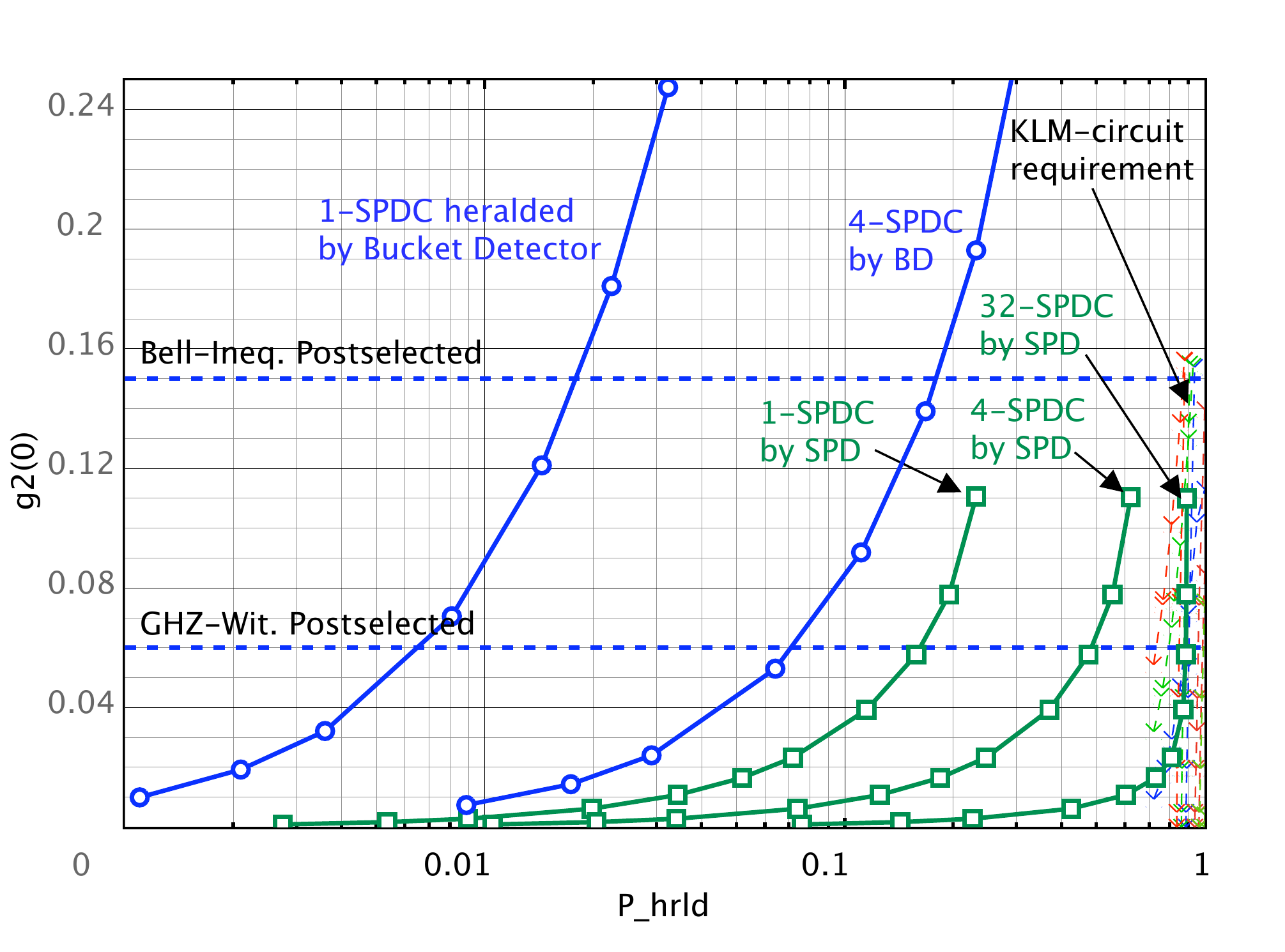}

\caption{Combination of the results from the numerical simulations of the LOQC with the single photon source from SPDC. The many superimposed plots in the right corner show the are of interest that is rqiured to be reached by a good single-photon source. The data named \emph{1-SPDC} corresponds to a photon source based on heralded photons created from spontaneous parametric down-conversion, heralded with either a bucket detector (BD) with efficiency $\eta=0.8$ or a single photon detector (SPD) with $\eta=0.95$. The switch for the SPDC photons  is a  binary switch, assuming each level has an transmittance of 0.98, and in addition, the coupling efficiency of the entangled photons is 0.98 each. The curve labled \emph{n-SPDC} shows the performance for photons generated from the switching of multiple SPDC sources. The parameter of the source curves is the squeezing parameter of the SPDC process. The faint plots on the lower right corner are the requirement data for achieving KLM-type operations outside the coincidence basis (see Figure~\ref{results}). In addition, the graph contains the thresholds for post-selective circuits such as creating a Bell-pair or a GHZ-triplet. In these cases the probability of the source, $P_{hrld}$ to generate a photon signal has no effect on the threshold. It is obvious, that photon sources based on heralding from one SPDC (1-SPDC) will never reach the areas of performance required for post-selection free LOQC operations. }
\label{sumarized_results}
\end{center}
\end{figure*}

Based on the SPDC state produced from squeezing, {\color{black}we want to study in detail the quality of heralded single photon states achieved by detecting one mode from the SPDC source, and correspondingly heralding a single photon in the other. Given the strict photon pair correlations imposed by the squeezing operator, this could deliver perfect single photons, } if it would not be for the higher order emissions. 
Various combinations and parameters of the heralded photons are modeled, and the produced state is analyzed. Since the detection process on the trigger photon corresponds, at least in principle, to a photon number measurement, the remaining photon is no longer in a superposition of number states, as originally given by the SPDC. Now it is actually a density matrix, which is represented as an array of number states weighted with amplitudes corresponding to the correct probability. In Figure \ref{sumarized_results}, the variation of {\color{black}$g^{(2)}(0)$} correlation vs the probability for heralding a photon are shown as the important characteristics of the created photon state, in dependance of the squeezing parameter $\epsilon$.

\begin{figure}[htbp]
\begin{center}
\includegraphics[width=13cm] {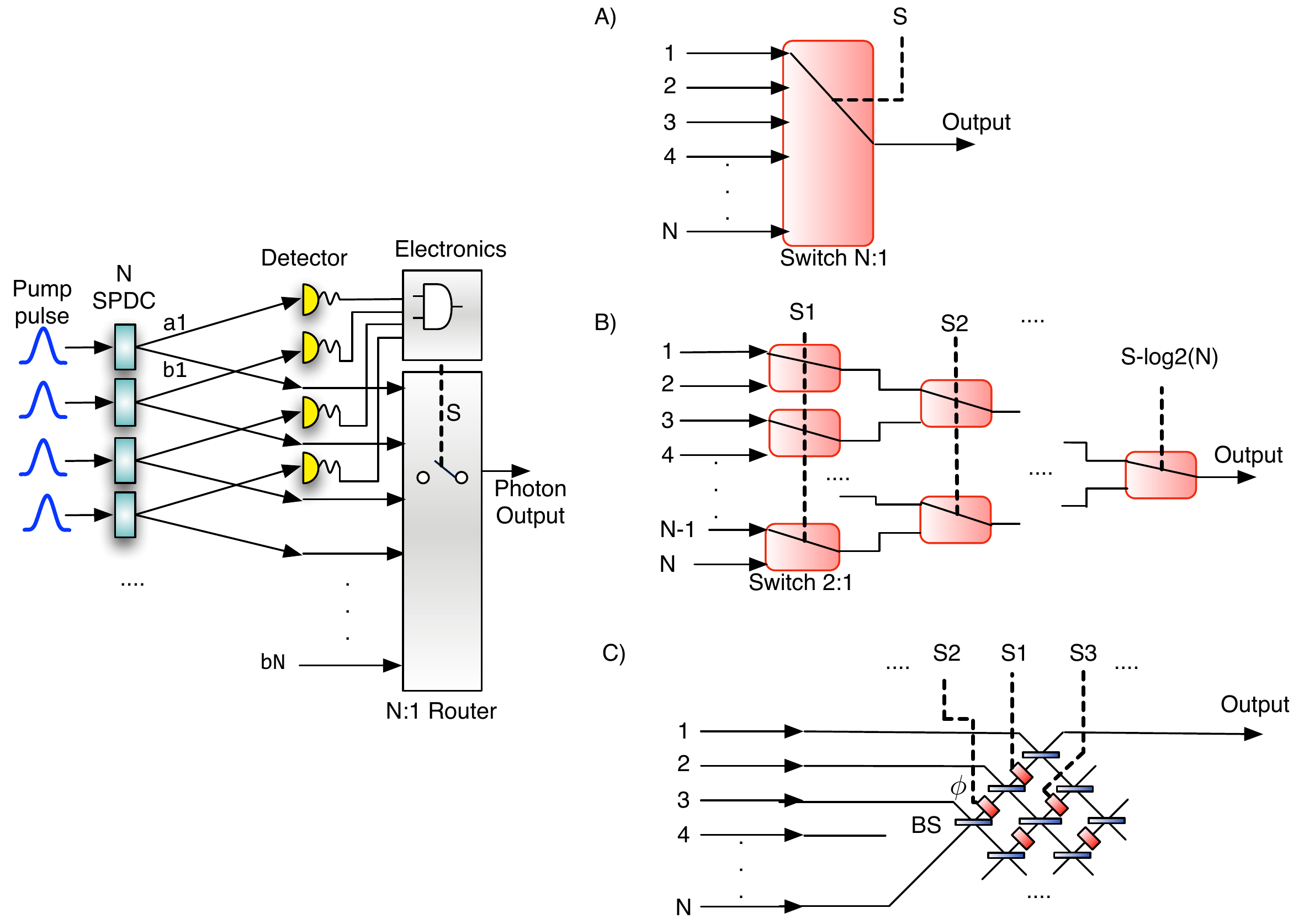}
\caption{(left) Heralding single photons from N spontaneous parametric down conversions (SPDC), each pumped by tightly synchronized ultra short pump pulses.  Depending on the trigger detection modes a1,...an, the corresponding signals from modes $b_1, ... b_N$ are switched to the output. (right) Depicted are three main architectures for implementing the routing switch. (a) one linear mode-switch, (b) binary coded mode-switches, (c) interferometric multi-port with phase modulators. The number of optical components in each photon path depends on the variant and therefore the loss will have fundamentally different scaling with the number of input modes N.}
\label{heralding_schemes}
\end{center}
\end{figure}


The output state from SPDC is generated from acting the operator on vacuum, $U_{SPDC} |0,0\rangle_{a_1,a_2}$, which yields:
\begin{equation}
|SPDC\rangle = C_0|00\rangle_{a_1,a_2} + C_1|11\rangle_{a_1,a_2} + C_2 |22\rangle_{a_1,a_2} + .... 
\end{equation}
the detection on the photon of side $a_1$ with a photon detector gives us the conditional output state on side $a_2$, leading to a mixture of photon number states, which can potentially be closer to an ideal single photon:
\begin{eqnarray}
\rho_{heralded 1} &=& Tr_{a1}(P_{BDa_1} |SPDC\rangle  )\\
&=&  \Bigl \{ Q |0\rangle \langle 0|_{a_2}; P_1 |C_1|^2  |1\rangle \langle 1|_{a_2}; P_2 |C_2|^2  |2\rangle \langle 2|_{a_2};  .... \Bigr \}
\end{eqnarray}
where $P_1, P_2,...$ are the detection probabilities for a "click" on the , as defined in expressions  on the trigger side for the respective photon number terms in Equation~\ref{proj_apd}, the $C_1, C_2,...$ are the amplitudes of the terms from the SPDC state, and $Q = 1 - \sum P_i |C_i|^2 $, is the probability of not finding a trigger detection event in mode $a_1$, or the vacuum term.  The  quality of one such a heralded single photon state is measured as the $g^{(2)}$-correlation function \cite{Mandel95a}, vs the probability for obtaining a successful heralding event. It is evident that heralded photons from SPDC are far from being useful for LOQC, as they are far from the required performance. In particular, the strength of the SPDC must be kept sufficient low such that the detection probability of the trigger is below 0.1 in order maintain a single photon purity of better than 90\%. (see the 1-SPDC plots in Figure~\ref{sumarized_results}). If the squeezing parameter, $\epsilon$ becomes to large, then the single-photon quality is drastically deteriorated.

The best approach in improving the output quality from from SPDC is to "chain" several such sources with active switches\cite{Kumar:2004fk, ma-2010}, see Figure \ref{heralding_schemes}. Thereby, several probabilistic sources are effectively combined into one deterministic source, which in principle can be made arbitrarily close to the ideal case. It is straightforward to study the characteristics of this switched SPDC source into the model, by keeping the squeezing parameter $\epsilon$ fixed at a certain value, and varying the number $S$ of SPDC sources that are combined. 
The heralded state is obtained from the original heralded single photon state, and yields 
\begin{equation}
\rho_{heralded S} = \Bigl \{ Q^S |0\rangle\langle0|_{a2};  \frac{(1-Q^S)}{(1-Q)}P_1 |C_1|^2 |1\rangle\langle1|_{a2};  \frac{(1-Q^S)}{(1-Q)}  P_2 |C_2|^2  |2\rangle\langle2|_{a2};  .... \Bigr \} ,   
\end{equation}
From the analysis shown in Figure \ref{sumarized_results} it is clear, that this type of heralding source outperforms a standard SPDC scheme, even at  modest squeezing parameters and number of sources of {\color{black}four}. In fact, this type of source may constitute a viable single photon source, given that other quantum systems such as quantum dots\cite{P.Michler12222000_b,AntonioBadolato05202005} or single stored ions \cite{1367-2630-11-10-103004} face the challenge of achieving good coupling efficiencies.

\begin{figure}[htbp]
\begin{center}
\includegraphics[width=8cm]{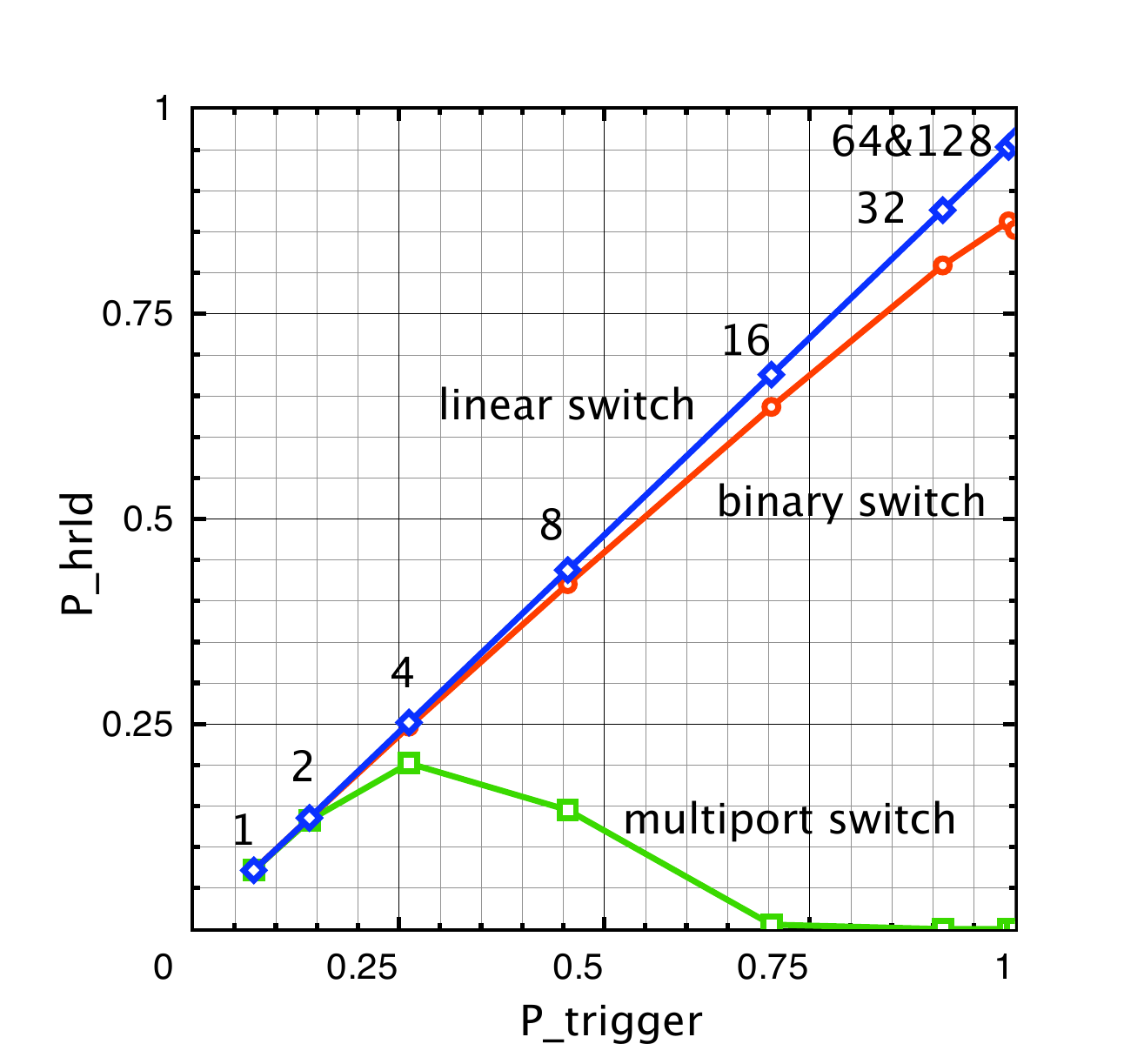}

\caption{Photon heralding efficiency vs the probability for a trigger, P\_trigger, for different numbers of sources (1,2,4, ... 128), as given as parameter on the plots.  P\_hrld is the probability of obtaining a non-zero photon output of the source. Three types a switches are considered, as discussed in the text, and the switching efficiency is assumed to be 0.98 per element.}
\label{heralding_efficiency}
\end{center}
\end{figure}

\subsection{Considering losses in the switches}

In any practical implementation of the SPDC-switch schemes the total losses will scale with the amount of input modes that should be accommodated. The schemes shown in  Figure \ref{heralding_schemes} each have their advantages and disadvantages, however it is crucial that they allow a scalable operation at all. Scheme A) is the most efficient, as each photon only passes on level of switches before entering the output mode. Therefore the overall system losses are constant, and don't change with the number of sources. Scheme B) is also very efficient, and consists of a cascade of binary switches. The amount of elements each photon must passage is on scales with the logarithm of the number of sources, $\log_2 (N)$.  Finally, scheme C) is a multi-input interferometer, with N inputs and 1 output, where the inputs are routed interferometrically to the output, e.g. via active phase changes. The interferometer follows the generalised rules for multi-ports\cite{Reck94b}, where on the order of $ N^2$ optical elements, including beam-splitters and phase shifters can realize any unitary operation.

{\color{black} From our simulation we obtained the probability for heralding a single-photon output vs trigger probability, for an increasing number of SPDC sources, {\color{black} see Figure~\ref{heralding_efficiency}.} These results clearly show that a linear (scheme A) or binary (scheme B) arrangement of switches can be useful for generating single-photon states sufficient for LOQC, whereas the multiport type (scheme C) of switch will fail due to the scaling of losses with additional switching layers. }

\section{Conclusions}
{\color{black} We described a numerical simulation of the quantum optics of linear optical quantum computing (LOQC), specifically tailored to obtain the characteristics of single-photon sources and photon detectors required to perform quantum computing operations in actual experiments.}

Firstly, we studied the heralding of Bell-pairs with scalable KLM-gates\cite{Knill:2001lr} where two photons, the control and target, are entangled via the assistance and detection of two ancillary photons.  These gates are fully scalable and could be combined into arbitrary quantum logic. The operation was verified by testing the correlations of the output photons against the CH inequality\cite{Clauser74a}, without post-selection.

Secondly, we simulated the creation of heralded Greenberger-Horne-Zeilinger (GHZ) states via fusing \cite{PhysRevLett.100.060502} two Bell-pairs created with separate KLM-gates  as the first building block of a cluster-state quantum computer\cite{raussendorf2001}. The resulting GHZ state was tested via a Mermin-type inequality\cite{Mermin90a} and a GHZ-state witness \cite{Guhne:2003fk}, all without post-selection.

Our results show that performing LOQC operations outside the post-selection basis requires photon-detectors with efficiencies of about $0.90$, single-photon sources with a success probabilities close to $0.90$ and single-photon purities ($g^{(2)}$-correlation function) better than $0.07$, as a guideline. These performance levels will be challenging to reach in real systems, but not impossible in our view. We compare these requirements with performance of state-of-the-art single photon sources based on down-conversion and cascaded down-conversion, as well other implementations. Our study is the first to establish precise guidelines for developing single-photon devices in respect of the implementing optical quantum processing applications.

\section{Acknowledgements}
{\color{black} We are grateful for useful discussions with T. Ralph, G. Milburn and T. Rudolph. } T.J. acknowledges support by the Australian Research Council (Linkage Fellowship). M.B. acknowledges support by the Marie Curie action of the EU Commission, {\color{black}PIEF-GA-2009-236345-PROMETEO.}

\pagebreak

\appendix
\section{Examples of states and operators represented in the numerical model}

Quantum states  are modelled as vectors, where each element is considered as the amplitude of this particular excitation. Therefore, all elements of the vectors and operators must obey the normalisation rules.

For example (for the dimension of the Fockspace  $N=4$), the vacuum state is a vector with amplitude on the first element

\begin{verbatim}
>> vacc
vacc = Quantum object
Hilbert space dimensions [ 4 ] by [ 1 ]
     1
     0
     0
     0
\end{verbatim}
Consequently, a single photon state is represented as the following vector:
\begin{verbatim}
oneph = Quantum object
Hilbert space dimensions [ 5 ] by [ 1 ]
     0
     1
     0
     0
     0	
\end{verbatim}
And the state that describes a superposition of  vacuum and one photon has the form:
\begin{verbatim}
>> 1/sqrt(2)*(vacc+oneph)
ans = Quantum object
Hilbert space dimensions [ 4 ] by [ 1 ]
      0.70711
      0.70711
            0
            0
\end{verbatim}

In this model, any operator acting on a single mode (such as the creation or annihilation operators, phase shifter,..) are represented as $N*N$ matrices.  Consequently, interaction Hamiltonians are expressed in the form of creation and annihilation operator matrices (example with $N=4$):
\begin{verbatim}
a = Quantum object
Hilbert space dimensions [ 4 ] by [ 4 ]
            0            1            0            0
            0            0       1.4142            0
            0            0            0       1.7321
            0            0            0            0
            
>> a'
ans = Quantum object
Hilbert space dimensions [ 4 ] by [ 4 ]
            0            0            0            0
            1            0            0            0
            0       1.4142            0            0
            0            0       1.7321            0
\end{verbatim}
and the unitary evolution is  obtained through exponentiation  $U_{bs} = expm(-i*H_{bs})$.

The extension ot a higher number of modes is realised by attaching the two (or more) modes thorugh the tensor product:
\begin{verbatim}
>> tensor(oneph,vacc)
ans = Quantum object
Hilbert space dimensions [ 5 5 ] by [ 1 1 ]
   (6,1)        1
\end{verbatim}
As an example, the operator representing a  50:50 beam splitter is a two mode operator is a $N^2 * N^2$ matrix. It
 is based on the two-mode Hamiltonian
\begin{equation}
H_{bs}=\theta(a^\dagger_1 a_2 + a_1 a^\dagger_2).
\end{equation}
As an example, this looks as follows:
\begin{verbatim}
%Beam splitter 50:50 = Quater wave plate @45¡
eta=1*pi/4; 
H_bs = (tensor(a,a') + tensor(a',a))*eta;
U_bs = expm(-1i*H_bs);

>> U_bs
U_bs = Quantum object
Hilbert space dimensions [ 5 5 ] by [ 5 5 ]
   (1,1)               1              
   (2,2)         0.70711              
   (6,2)               0 -    0.70711i
   (3,3)             0.5              
   (7,3)               0 -    0.70711i
  (11,3)            -0.5              
   (4,4)         0.35355              
   (8,4)               0 -    0.61237i
  (12,4)        -0.61237              
  (16,4)               0 +    0.35355i
   (5,5)            0.25              
   (9,5)               0 -        0.5i
   ............
\end{verbatim}

Density matrices are represented also as matrices, and also can be subject to tensor-products, evolution with operators as well as detections.  For example, the single photon state will have a density operator of the form:

The actual matrices representing the detection are determined according to equations~\ref{proj_apd} and \ref{proj_spd}. For the example of a bucket-detector in a Fock-space the of dimensionality $N=4$, an efficiency $\eta=0.6$, and a noise probability $ \epsilon_d= 0.001$, the detector matrices take the form:
\begin{verbatim}
>> oneph*oneph'
ans = Quantum object
Hilbert space dimensions [ 4 ] by [ 4 ]
     0     0     0     0
     0     1     0     0
     0     0     0     0
     0     0     0     0.
\end{verbatim}
And as for the example of the equal superposition of the vacuum state and one photon, the density matrix takes the form:
\begin{verbatim}
>> sup=1/sqrt(2)*(vacc+oneph);
>> sup*sup'
ans = Quantum object
Hilbert space dimensions [ 4 ] by [ 4 ]
          0.5          0.5            0            0
          0.5          0.5            0            0
            0            0            0            0
            0            0            0            0.
\end{verbatim}

This can also be represented in the form of Ket and Bra, 
\begin{verbatim}
>> display_state4(N,sup)

ans = 

    'Pure state=
 0.707 * '
    '|0,> 1.000 '
    '|1,> 1.000 '
\end{verbatim}

\section{Single-Photon Detectors}

We give examples of the concrete representation of single photon detectors. The operator for a Bucket-detector, with efficiency of $0.6$, a noise factor of $0.001$ has the form:

\begin{verbatim}
>> [bd bd_n]=BucketDetector_noise(N,0.6,0.001)
bd = Quantum object
Hilbert space dimensions [ 4 ] by [ 4 ]
        0.001            0            0            0
            0        0.601            0            0
            0            0        0.841            0
            0            0            0        0.937

bd_n = Quantum object
Hilbert space dimensions [ 4 ] by [ 4 ]
        0.999            0            0            0
            0        0.399            0            0
            0            0        0.159            0
            0            0            0        0.063
\end{verbatim}
where $bd$ is used to calculate the probability for obtaining a detection event, and $bd_n$ the probability for not obtaining a detection. The actual probabilities for the detection of a photon, are obtained by taking the expectation value of the detector matrix with the respected quantum state. 

For our example with the previously defined operators  $bd$ and  $bd_n$, the actual probabilities for observing a Click or a Non-Click for a one-photon state, {\tt oneph}, are easily computed as expectation values:
\begin{verbatim}
>> (oneph'*bd*oneph)
ans = Quantum object
Hilbert space dimensions [ 1 ] by [ 1 ]
        0.601

>> (oneph'*bd_n*oneph)
ans = Quantum object
Hilbert space dimensions [ 1 ] by [ 1 ]
        0.399.
\end{verbatim}

\section{Modelling spontaneous parametric down-conversion}

  The evolution of the Hamiltonian in Eq.~\ref{H_spcd} which describes SPDC,  can be calculated via matrix exponentiation, $U_{SPDC} = expm (-i H)$, which for example takes the form ( $N=4$, $\epsilon=0.4$) :
\begin{verbatim}

%SPDC in chi2:
H_chi2=(tensor(a,a)+tensor(a',a'))*epsilon;
U_chi2=expm(-1i*H_chi2);

%SPDC input state for pair of photons in HH
spdc_state=U_chi2*tensor(vacc,vacc);

U_chi2 = Quantum object
Hilbert space dimensions [ 4 4 ] by [ 4 4 ]
   (1,1)           0.925              
   (6,1)               0 -    0.35162i
  (11,1)        -0.13218              
  (16,1)               0 +   0.057186i
   (2,2)         0.85635              
   (7,2)               0 -     0.4525i
  (12,2)         -0.2488              
   (3,3)         0.76945              
   (8,3)               0 -    0.63871i
   (4,4)               1              
   ...
  \end{verbatim}

 It is straight forward  to calculate the state produced by the SPDC operator, by multiplying this operator matrix with a given input state, in this case vaccum (i.e. $|0\rangle =  [1,0,0,0]$), $|SPDC\rangle = U_{SPDC} * tensor([1,0,0,0],[1,0,0,0])$, giving the output state of the down-converter:
  \begin{verbatim}
  spdc_state = Quantum object
Hilbert space dimensions [ 4 4 ] by [ 1 1 ]
   (1,1)           0.925              
   (6,1)               0 -    0.35162i
  (11,1)        -0.13218              
  (16,1)               0 +   0.057186i
    \end{verbatim}
  where the (1,1) term corresponds to $|00\rangle_{a_1,a_2}$ output, the (6,1) term corresponds to $|11\rangle_{a_1,a_2}$, (11,1) corresponds to $|22\rangle_{a_1,a_2}$ and so on. Expressed with Kets representing the photon number, the SPDC state has the following form:
  \begin{verbatim}
  >> display_state4(N,spdc_state)

ans = 

    'Pure state=
 0.925 * '
    '|0,0,> 1.000 '
    '|1,1,> 0.000-0.380i '
    '|2,2,> -0.143 '
    '|3,3,> 0.000+0.062i '.
\end{verbatim}


\begin{thebibliography}{26}
\providecommand{\natexlab}[1]{#1}

\bibitem[1]{Knill:2001lr}
Knill, E.; Laflamme, R.; Milburn, G.J. A scheme for efficient quantum
  computation with linear optics.  {\em Nature}  {\bf 2001}, {\em 409} (6816),
  46--52.

\bibitem[2]{raussendorf2001}
Raussendorf, R.; Briegel, H.J. A One-Way Quantum Computer.  {\em Phys. Rev.
  Lett.}  {\bf 2001}, {\em 86} (22) (May), 5188--5191.

\bibitem[3]{browne-2005-95}
Browne, D.E.; Rudolph, T. Resource-efficient linear optical quantum
  computation.  {\em Physical Review Letters}  {\bf 2005}, {\em 95}, 010501.

\bibitem[4]{PhysRevLett.100.060502}
Varnava, M.; Browne, D.E.; Rudolph, T. How Good Must Single Photon Sources and
  Detectors Be for Efficient Linear Optical Quantum Computation?.  {\em Phys.
  Rev. Lett.}  {\bf 2008}, {\em 100} (6) (Feb), 060502.

\bibitem[5]{PhysRevA.81.052303}
Gong, Y.X.; Zou, X.B.; Ralph, T.C.; Zhu, S.N.; et~al. Linear optical quantum
  computation with imperfect entangled photon-pair sources and inefficient
  non--photon-number-resolving detectors.  {\em Phys. Rev. A}  {\bf 2010}, {\em
  81} (5) (May), 052303.

\bibitem[6]{1464-4266-1-4-312}
Tan, S.M. A computational toolbox for quantum and atomic optics.  {\em Journal
  of Optics B: Quantum and Semiclassical Optics}  {\bf 1999}, {\em 1} (4),
  424--432.

\bibitem[7]{Jennewein2010}
Jennewein, T. Toolbox for Quantum Photonics in Matlab.  {\em Download:
  http://info.iqc.ca/qpl/}  {\bf 2010}.

\bibitem[8]{Bouwmeester97a}
Bouwmeester, D.; Pan, J.W.; Mattle, K.; Eibl, M.; Weinfurter, H.; et~al.
  Experimental Quantum Teleportation.  {\em Nature}  {\bf 1997}, {\em 390},
  575--579.

\bibitem[9]{Zukowski95a}
Zukowski, M.; Zeilinger, A.; Weinfurter, H. Entangling photons radiated by
  independent pulsed sources.  {\em Annals of the N.Y. Acad. of Sciences}  {\bf
  1995}, {\em 755}, 91--102.

\bibitem[10]{PhysRevLett.95.210504}
Langford, N.K.; Weinhold, T.J.; Prevedel, R.; Resch, K.J.; Gilchrist, A.;
  O\char39{}Brien, J.L.; Pryde, G.J.; et~al. Demonstration of a Simple
  Entangling Optical Gate and Its Use in Bell-State Analysis.  {\em Phys. Rev.
  Lett.}  {\bf 2005}, {\em 95} (21) (Nov), 210504.

\bibitem[11]{PhysRevLett.91.227901}
Barbieri, M.; De~Martini, F.; Di~Nepi, G.; Mataloni, P.; D'Ariano, G.M.; et~al.
  Detection of Entanglement with Polarized Photons: Experimental Realization of
  an Entanglement Witness.  {\em Phys. Rev. Lett.}  {\bf 2003}, {\em 91} (22)
  (Nov), 227901.

\bibitem[12]{PhysRevLett.23.880}
Clauser, J.F.; Horne, M.A.; Shimony, A.; et~al. Proposed Experiment to Test
  Local Hidden-Variable Theories.  {\em Phys. Rev. Lett.}  {\bf 1969}, {\em 23}
  (15) (Oct), 880--884.

\bibitem[13]{Guhne:2003fk}
G{\"u}hne, O.; Hyllus, P. Investigating Three Qubit Entanglement with Local
  Measurements.  {\em International Journal of Theoretical Physics}  {\bf
  2003}, {\em 42} (5) (05), 1001--1013.

\bibitem[14]{PhysRevA.68.042328}
Dodd, J.L.; Ralph, T.C.; Milburn, G.J. Experimental requirements for
  Grover\char39{}s algorithm in optical quantum computation.  {\em Phys. Rev.
  A}  {\bf 2003}, {\em 68} (4) (Oct), 042328.

\bibitem[15]{PhysRevA.66.052306}
Knill, E. Quantum gates using linear optics and postselection.  {\em Phys. Rev.
  A}  {\bf 2002}, {\em 66} (5) (Nov), 052306.

\bibitem[16]{PhysRevA.64.062311}
Pittman, T.B.; Jacobs, B.C.; Franson, J.D. Probabilistic quantum logic
  operations using polarizing beam splitters.  {\em Phys. Rev. A}  {\bf 2001},
  {\em 64} (6) (Nov), 062311.

\bibitem[17]{Clauser74a}
Clauser, J.F.; Horne, M.A. Experimental consequences of objective local
  theories.  {\em Phys. Rev.~D}  {\bf 1974}, {\em 10} (2), 526--535.

\bibitem[18]{Mermin90a}
Mermin, N.D. Extreme quantum entanglement in a superposition of macroscopically
  different states.  {\em Phys. Rev. Lett.}  {\bf 1990}, {\em 65} (15),
  1838--1840.

\bibitem[19]{N.:1967kb}
Klyshko, D.N. Coherent Photon Decay in a Nonlinear Medium.  {\em JETP Lett.}
  {\bf 1967}, {\em 6} (1), 23.

\bibitem[20]{Mandel95a}
Mandel, L.; Wolf, E. Optical Coherence and Quantum Optics.  Cambridge
  University Press: Cambridge, 1995.

\bibitem[21]{Kumar:2004fk}
Kumar, P.; Kwiat, P.; Migdall, A.; Nam, S.W.; Vuckovic, J.; et~al. Photonic
  Technologies for Quantum Information Processing.  {\em Quantum Information
  Processing}  {\bf 2004}, {\em 3} (1) (10), 215--231.

\bibitem[22]{ma-2010}
Ma, X.S.; Zotter, S.; Kofler, J.; Jennewein, T.; et~al., Towards on-demand
  single-photon generation via active multiplexing, , 2010.

\bibitem[23]{P.Michler12222000_b}
Michler, P.; Kiraz, A.; Becher, C.; Schoenfeld, W.V.; Petroff, P.M.; Zhang, L.;
  Hu, E.; et~al. {A Quantum Dot Single-Photon Turnstile Device}.  {\em Science}
   {\bf 2000}, {\em 290} (5500), 2282--2285.

\bibitem[24]{AntonioBadolato05202005}
Badolato, A.; Hennessy, K.; Atature, M.; Dreiser, J.; Hu, E.; Petroff, P.M.;
  et~al. {Deterministic Coupling of Single Quantum Dots to Single Nanocavity
  Modes}.  {\em Science}  {\bf 2005}, {\em 308} (5725), 1158--1161.

\bibitem[25]{1367-2630-11-10-103004}
Barros, H.G.; Stute, A.; Northup, T.E.; Russo, C.; Schmidt, P.O.; et~al.
  Deterministic single-photon source from a single ion.  {\em New Journal of
  Physics}  {\bf 2009}, {\em 11} (10), 103004.

\bibitem[26]{Reck94b}
Reck, M.; Zeilinger, A.; Bernstein, H.J.; et~al. Experimental realization of
  any discrete unitary operator.  {\em Phys. Rev. Lett.}  {\bf 1994}, {\em 73}
  (1), 58--61.

\end{thebibliography}
\end{document}